%% file: 0main.tex
\documentclass[sigconf,authorversion,nonacm]{acmart}
\usepackage{multirow}
\usepackage{algorithm}
\usepackage{algorithmicx}
\usepackage{algpseudocode}
\usepackage{amsmath}
\usepackage{mathtools}
\usepackage{amsthm}
\usepackage{enumitem}

\newcommand{\eg}{\textit{e}.\textit{g}., }
\newcommand{\ie}{\textit{i}.\textit{e}., }

\newtheorem{remark}{Remark}

\newtheorem{proposition}{Proposition} 

\newcounter{bxincomm}
\definecolor{aqua}{rgb}{0.00,0.67,0.80}

\newcounter{todocomm}

\newcommand{\model}{\textsc{IGPrune}}
\input{math_commands.tex}

\AtBeginDocument{%
  }

\setcopyright{acmlicensed}
\copyrightyear{2026}
\acmYear{2026}
\acmDOI{XXXXXXX.XXXXXXX}
\acmConference[Conference acronym 'XX]{Submission to WWW 2025}{June 03--05,
  2018}{Woodstock, NY}
\acmISBN{978-1-4503-XXXX-X/2018/06}

\begin{document}

\title{Preserving Core Structures of Social Networks\\via Information Guided Multi-Step Graph Pruning}


\author{Yutong Hu}
\authornote{These authors contributed equally to this work.}
\affiliation{%
  \institution{Shanghai Jiao Tong University}
  \city{Shanghai}
  \country{China}
}

\author{Bingxin Zhou}
\authornotemark[1]
\authornote{Corresponding authors (bingxin.zhou@sjtu.edu.cn, hongl3liang@sjtu.edu.cn).}
\affiliation{%
  \institution{Shanghai Jiao Tong University}
  \city{Shanghai}
  \country{China}
}

\author{Jing Wang}
\affiliation{%
  \institution{Shanghai Jiao Tong University}
  \city{Shanghai}
  \country{China}
}

\author{Weishu Zhao}
\affiliation{%
  \institution{Shanghai Jiao Tong University}
  \city{Shanghai}
  \country{China}
}

\author{Liang Hong}
\authornotemark[2]
\affiliation{%
  \institution{Shanghai Jiao Tong University}
  \city{Shanghai}
  \country{China}
}

\renewcommand{\shortauthors}{Hu and Zhou et al.}

\begin{abstract}
Social networks often contain dense and overlapping connections that obscure their essential interaction patterns, making analysis and interpretation challenging. Identifying the structural backbone of such networks is crucial for understanding community organization, information flow, and functional relationships. This study introduces a multi-step network pruning framework that leverages principles from information theory to balance structural complexity and task-relevant information. The framework iteratively evaluates and removes edges from the graph based on their contribution to task-relevant mutual information, producing a trajectory of network simplification that preserves most of the inherent semantics. Motivated by gradient boosting, we propose \model, which enables efficient, differentiable optimization to progressively uncover semantically meaningful connections. Extensive experiments on social and biological networks show that \model~retains critical structural and functional patterns. Beyond quantitative performance, the pruned networks reveal interpretable backbones, highlighting the method’s potential to support scientific discovery and actionable insights in real-world networks. 
\end{abstract}




\maketitle

\section{Introduction}
Graphs are widely used to represent complex systems, from social and biological networks to citation and knowledge graphs \citep{ye2022survey, sharma2024survey, khemani2024survey, khoshraftar2024survey} . A key challenge is that raw graphs often contain an excessive number of links, while the system’s essential organization can be captured by only a subset of them \citep{ahnert2014generalised}. As a result, important patterns are frequently buried under incidental connections unless the semantic backbone of the graph is extracted. For instance, in social networks, the backbone is formed by strong ties and bridging connections between communities \citep{tabassum2018social, doreian2012social}. In gene co-occurrence networks, the semantic backbone corresponds to central cycles like carbon and nitrogen metabolism over other pathways \citep{petralia2015integrative, vignes2011gene}. In citation networks, it highlights seminal works and influential cross-disciplinary references that structure the flow of knowledge \citep{mclaren2022citation, zhao2015analysis}. Identifying such backbones is thus essential for uncovering the core patterns of a system of interactions.

Building on this motivation, a common approach is to simplify graphs before further analysis. Two main strategies have been developed. (1) \emph{Graph summarization} \citep{kang2022personalized} reconstructs the original graph into a coarser representation by grouping nodes into clusters, which improves computational efficiency and visualization. However, by merging inter-cluster entities and removing fine-grained connections, this approach loses the ability to perform detailed, node-level analysis. (2) \emph{Edge sparsification} \citep{hashemi2024comprehensive} reduces the number of edges based on heuristics or task-specific criteria. While it can eliminate redundant links, the contribution of the retained edges to the network’s organization is often unclear, which lead to the potential removal of semantically important edges. Both approaches focus primarily on computational scalability and memory efficiency, leaving the explicit inference of the simplified graph underexplored.

This study adopts an information-theoretic perspective on graph pruning to address these identified limitations. We define \emph{mutual information} (MI) as a measure of graph semantics. It quantifies the graph’s structural information remained after removing edges. MI has been previously deployed to evaluate the information content of representations \citep{zou2024ib, furutanpey2025adversarial, shwartz2017opening}. Based on this measure, we introduce the local \textit{Information–Complexity score} and associated global measurements. They serve as unified, task-agnostic criteria for evaluating the quality of graph pruning through the trade-off between preserving informative structures and reducing redundancy.

\begin{figure*}
    \centering
    \includegraphics[width=0.85\linewidth]{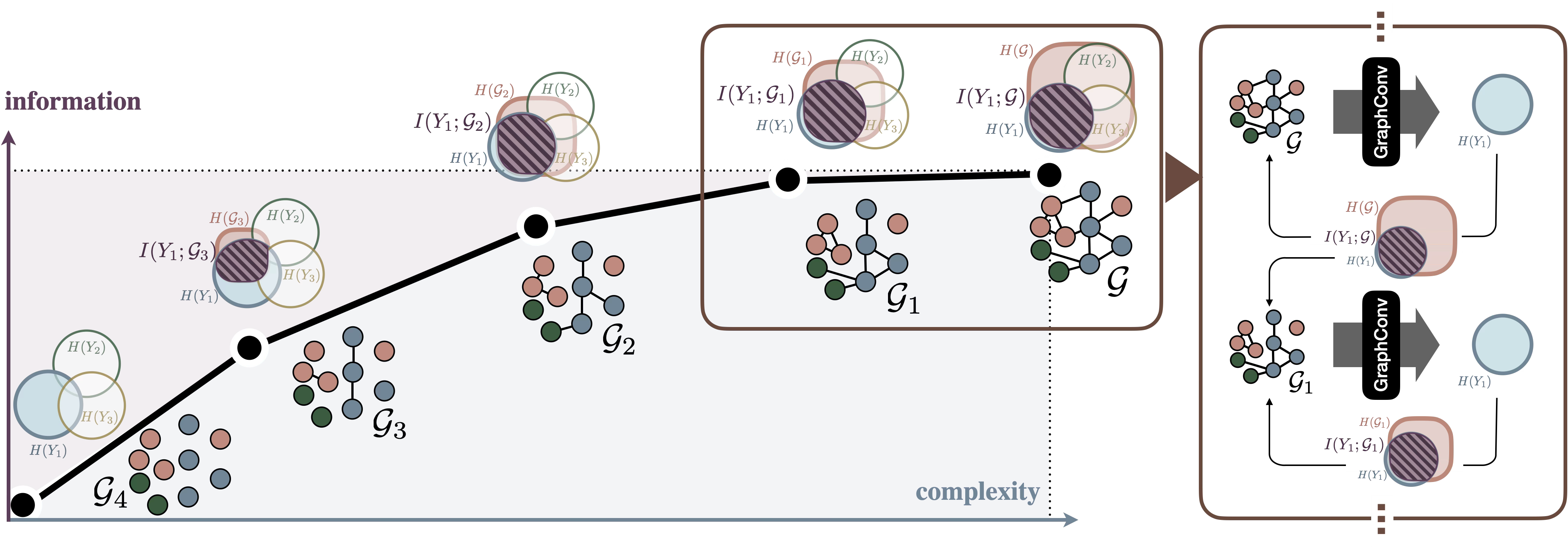}
    \caption{Illustrative Figure. \model~iteratively prunes edges based on gradients guided by information $I(Y_1;\gG_3)$, gradually focusing the graph representation from raw structural pattern $H(\gG)$ to task-relevant information $H(Y_1)$.}
    \label{fig:architecture}
\end{figure*}

We design \model, a gradient-guided pruning framework (Figure~\ref{fig:architecture}). Inspired by the iterative refinement principle in gradient boosting, \model~progressively updates edge weights according to the gradient of the objective. Each step removes edges contributing least to MI, and the graph representation is updated accordingly. This stepwise process enables the model to reduce structural redundancy while maintaining task-relevant information flow.

To validate the effectiveness and interpretability of \model, we conduct both quantitative and qualitative studies on diverse social and biological networks. Extensive results show that \model~preserves structural information even under high sparsity. Moreover, when applied to microbial gene co-occurrence networks from opposite environmental extremes, \model~uncovers environment-specific graph semantics of metabolic pathways from different systems, which demonstrates the practical utility of the pruned graph trajectory for scientific discovery.

In summary, this work makes four key contributions. First, we define a new graph pruning task that seeks an informative and interpretable pruning trajectory, preserving semantic structures while reducing complexity. Second, we propose differentiable optimization objectives for MI maximization, enabling gradient-based optimization under the empirical risk minimization framework. Third, we establish theoretical links between graph pruning, information theory, and graph semantics, providing an information-theoretic foundation for structural simplification. Finally, we demonstrate the practical utility of the proposed framework through real-world applications, showing its ability to extract interpretable graph backbones that facilitate scientific discovery.

\section{Related Work}
\paragraph{Graph Summarization}
Graph summarization has been studied from both task-specific and general-purpose perspectives.
Task-specific methods focus on particular application domains, while general approaches seek domain-agnostic techniques for structural simplification. Common strategies include graph reconstruction \citep{kumar2018utility,singh2021utility}, clustering \citep{boobalan2016graph}, pattern mining \citep{liu2008summarizing}, and sampling \citep{beg2018scalable,navlakha2008graph}. Representative works include spectral sparsification \citep{spielman2011spectral}, which approximates graph Laplacians with sparse subgraphs,; backbone extraction \citep{serrano2009backbone} identifies statistically significant edges; Pattern- and MDL-based methods (\eg \textsc{VoG} \citep{koutra2015vog}) compress graphs into interpretable motifs \citep{liu2018graph}. Beyond sparsification, coarsening and condensation \citep{hashemi2024comprehensive} produce smaller proxy graphs by merging nodes or learning task-preserving condensed graphs. \textsc{PRI-Graphs} \citep{yu2022principle} extend the Principle of Relevant Information to graphs, leveraging Laplacian structures for information-preserving sparsification. In graph neural networks (GNNs), pruning techniques~\citep{jin2021graph,wang2022pruning} remove edges or nodes to improve efficiency.

\paragraph{Edge Sparsification}
Edge sparsification compresses graph information by selectively preserving or removing edges while keeping the node set unchanged. The goal is to retain global or local properties (\eg distances, cuts, or spectral characteristics) or maintain downstream task performance. In classical graph theory, a subgraph preserving pairwise distances is a \emph{spanner} \cite{peleg1989graph}, while one preserving cut or spectral structures is a \emph{sparsifier} \cite{spielman2011spectral}. These concepts establish the theoretical foundation for modern sparsification techniques applied to large-scale graphs. Traditional methods use heuristic or combinatorial criteria, \eg edge betweenness, effective resistance, and spectral similarity, for edge reduction. These task-agnostic approaches may ignore semantic dependencies. Recent works integrate sparsification with GNNs, performing task-aware edge pruning via gradient-based importance \citep{lee2018snip,wang2020picking,zhou2023robust} or learned structural masks \citep{chen2021graph,zhou2024graph}. 
Despite their success, these methods emphasize accuracy preservation rather than interpretability or stability. 

\paragraph{Mutual Information in Deep Learning}
MI has long guided deep learning for balancing compression with task-relevant fidelity. Classical work such as the Information Bottleneck (IB) framework \citep{shwartz2017opening, shwartz2022information}. A recent graph-specific example is \textsc{PRI-Graphs} \citep{yu2022principle}, which extends the \emph{Principle of Relevant Information} to graphs by selecting edges to preserve structural information via the graph Laplacian. Other works share the underlying MI idea though applied in different domains. \textsc{IB-AdCSCNet} \citep{zou2024ib} uses IB to guide sparse coding in convolutional networks for image classification, and bottleneck-injected DNNs for task-oriented communication \citep{furutanpey2025adversarial} invoke IB objectives to control information flow and robustness under communication constraints. While these methods are not designed for graph edge pruning or summarization, hey demonstrate that MI-based trade-offs between information retention and complexity are broadly useful and motivate our multi-step pruning approach.


\section{Problem Formulation: Graph Semantics as Mutual Information}
This section formalizes the graph pruning problem and introduces the information-theoretic view. Section~\ref{sec:setup} defines basic notations and graph semantics. Section~\ref{sec:kstep} describes multi-step pruning as a structured simplification process, and Section~\ref{sec:MI} links it to information theory via MI. Finally, Section~\ref{sec:MI_estimate} presents a classifier-based approach to estimate MI.

\subsection{Problem Setup and Notations}
\label{sec:setup}
Consider a graph $\gG(\gV,\gE)$ with $|\gV|=n$ nodes and $|\gE|=E$ edges, where $\mA \in \mathbb{R}^{n \times n}$ is the adjacency matrix and $\mX \in \R^{n \times d}$ contains $d$-dimensional node features.

\begin{definition}[Graph Semantics]
Let $\gG(\gV,\gE)$ be a graph with adjacency matrix $\mA$ and node features $\mX$. The \emph{graph semantics} refers to a subset of nodes $\gV^{\prime}\subseteq\gV$ and edges 
$\gE^{\prime}\subseteq\gE$ that preserves the essential structural information of the original graph. 
\end{definition}

Intuitively, graph semantics corresponds to the backbone of the input graph that maintains the organization and dependencies most relevant to understanding the system. In later sections, this notion will be quantified through MI (Section~\ref{sec:MI}). 

\subsection{Multi-Step Graph Pruning}
\label{sec:kstep}
We formulate graph pruning as a sequential process of $K$ steps that progressively simplify the adjacency matrix $\{\mathbf{A}_0, \mathbf{A}_1, \dots, \mathbf{A}_K\}$, where $\mathbf{A}_0 = \mathbf{A}$.  
At each step $k$, a subset of edges is removed from $\mathbf{A}_{k-1}$ according to a pruning operator $\gP_k(\cdot)$, \ie
\begin{equation}
    \mathbf{A}_k = \gP_k(\mathbf{A}_{k-1}).
\end{equation}
Each operator $\gP_k(\cdot)$ removes up to $\tau_k$ edges, where $\tau_k$ denotes the pruning budget at step $k$. When the pruning process is divided into $K$ steps, each step removes an equal fraction of $E$ edges, \ie $\tau_k = {E}/{K}$. This yields a graph sequence $\gG \to \gG_1 \to \cdots \to \gG_K$, where each $\gG_k$ depends only on its predecessor. This is analogous to a Markov chain, which enables stepwise analysis of information degradation during pruning. Formally,
\begin{equation}
    P(\gG_{k+1} \mid \gG_k, \gG_{k-1},\dots,\gG_1) = P(\gG_{k+1} \mid \gG_k).
\end{equation}
This property allows us to track how information is lost step by step without having to consider the entire pruning history.

\subsection{Mutual Information for Graph Pruning}
\label{sec:MI}
Recall the definition of MI between random variables $X$ and $Z$:
\begin{equation}
\label{eq:mi_def}
    I(X;Z) \coloneq H(X) - H(X\mid Z),
\end{equation}
where $H(\cdot)$ denotes Shannon entropy \citep{shannon1948mathematical}. Here, $I(X;Z)$ quantifies the reduction in uncertainty about $X$ when $Z$ is observed, \ie the amount of information $Z$ provides about $X$. 

This formulation allows measuring the information shared between two random variables, which motivates us to assess the information preserved by a simplified graph $\gG_k$ relative to the original $\gG$ in graph pruning. We consider information preservation from two complementary perspectives of \emph{structural information} and \emph{task-specific information} (\eg node-level labels), respectively.

\paragraph{Structural information preservation}  
The first view measures the retention of $\gG_k$ with respect to the organization of the original graph $\gG$. This is a typical objective considered in graph summarization \citep{kumar2018utility}. At the $k$th step, the MI between $\gG$ and $\gG_k$ is $I(\gG;\gG_{k})$. Since removing edges do not increase information, it holds that
\begin{equation}
    0 \le I(\gG;\gG_k) \le I(\gG;\gG_{k-1}) \le H(\gG).
\end{equation} 
The closer $I(\gG;\gG_k)$ is to $H(\gG)$, the more structural information is preserved in the simplified graph.

\paragraph{Task-relevant information}  
The second view focuses on the information relevant to a downstream node-level task. Let $\mY$ denote the task level, then $I(\mY;\gG_k)$ measures how much information about $\mY$ is retained after step $k$. By data processing inequality (DPI) \citep{shwartz2017opening}, MI decreases monotonically during pruning, \ie
\begin{equation}
    I(\mY;\gG) \ge I(\mY;\gG_1) \ge \cdots \ge I(\mY;\gG_K).
\end{equation}
Thus, pruning preserves task-critical edges while removing irrelevant ones. This aligns the simplification process with practical objectives such as node classification accuracy. Accordingly, we set our objective of graph pruning as monitoring the information loss so that it preserves as much task-relevant information as possible.

\subsection{Task-Based Mutual Information Estimation}
\label{sec:MI_estimate}
Although MI offers a rigorous pruning criterion, the unknown joint distribution $P(\gG,\mY)$ hinders direct optimization of $I(\mY;\gG_k)$. We therefore approximate $p(\mY\mid\gG)$ using a learnable predictor (\eg, GNNs). This formulation applies to both classification and regression tasks, as the conditional entropy $H(\mY\mid\gG)$ can be estimated from any parametric likelihood model, regardless of output type.

\begin{proposition}[Predictor-Based Mutual Information Lower Bound]
\label{prop:MI_est}
Let $q_\phi(\mY\mid\gG)$ be a parametric predictor with learnable parameters $\phi$, trained on labeled samples $\{(\gG_i, y_i)\}_{i=1}^M$ to approximate $p(\mY\mid\gG)$. 
Then, the MI between the graph $\gG$ and the task target $\mY$ can be lower-bounded by the empirical negative log-likelihood (NLL) of the predictor:
\begin{equation}
\label{eq:I_estimate}
    \widehat{I}_{q_\phi}(\gG;\mY)
    \approx 
    \hat{H}(\mY)
    - \frac{1}{M}\sum_{i=1}^M [-\log q_\phi(y_i \mid \gG_i)],
\end{equation}
where $M$ denotes the number of labeled samples, $y_i$ is the task label of sample $i$, and $\hat{H}(\mY)$ is the empirical entropy of the target variable.
\end{proposition}

The proof of Proposition~\ref{prop:MI_est} is provided in Appendix~\ref{app:proof}. In practice, (\ref{eq:I_estimate}) can also be applied to a hold-out set to evaluate the task-relevant MI of a pruned graph. 
Specifically, given a set of unseen samples $\{(\gG_i, y_i)\}_{i=1}^{M_\text{hold}}$, we have
\begin{equation}\label{eq:holdout_MI_est}
    \widehat{I}_{q_\phi}(\gG;\mY)_\text{hold}
    \approx
    \hat{H}(\mY_\text{hold})
    - \frac{1}{M_\text{hold}}\sum_{i=1}^{M_\text{hold}} [-\log q_\phi(y_i\mid\gG_i)],
\end{equation}
where $\hat{H}(\mY_\text{hold})$ is the empirical entropy of the hold-out labels. 
This provides a practical metric to quantify how much task-relevant information is retained in a pruned graph, without requiring access to the true distribution $p(\mY\mid\gG)$.

The above proposition directly shows the consistency between minimizing the empirical loss of a node-level predictor and maximizing the lower bound of task-relevant MI. The bound is tight when the predictor exactly matches the true conditional distribution, and the gap corresponds to the expected KL divergence. Hence, \textbf{improving empirical predictive performance inherently increases the estimated task-relevant MI}. This provides a tractable objective for evaluating information preserved in pruned graphs.

\begin{remark}
Since $\widehat{I}_{q_\phi}(\gG;\mY)$ is a lower bound of the true MI, the measured information in experiments may not strictly decrease across pruning steps. 
Nonetheless, this estimator remains a meaningful metric for assessing task-relevant information retention.
\end{remark}

\section{Evaluation Metrics: Local and Global Scoring}
\label{sec:IC_score}
Graph pruning simplifies edge connections while retaining task-relevant information. Evaluation occurs at two levels: locally, the \emph{complexity} $\mC$ and \emph{information} $\mI$ scores quantify structural reduction and predictive information at each step $k$; globally, these scores are aggregated over $K$ steps into metrics such as the \emph{Area Under the Information–Complexity Curve} and the \emph{Information-Budget Point}, which together capture the trade-off between simplification and information retention for comprehensive comparison.

\subsection{Local Measurements}
We first define the fundamental scores for each pruning step $k$, \ie the complexity score and the information score.

\begin{definition}[Complexity Score]
In a stepwise graph pruning process $\gG \to \gG_1 \to \cdots \to \gG_K$, the complexity score of a pruned graph $\gG_k$ with adjacency matrix $\mA_k$ is
\begin{equation}
    \gC(\mA_k) \coloneqq \frac{\gE(\mA_k)}{\gE(\mA_0)},
\end{equation}
where $\gE(\mA_k)$ is a structural energy measure.
\end{definition}
By this definition, $\mC(\mA_0)=1$ for the original graph and $\mC(\mA_K)=0$ for the final edgeless graph $\gG_K$. In this study, we use $\gE(\mA_k)=|\mA_k|_0$, the number of nonzero edges, in experiments. In practice, the structural energy measure can also be defined from other perspectives, such as the cut size and motif counts. 
Alternatively, other structural energy measures can be employed depending on the application, such as cut size or motif counts.

\begin{definition}[Information Score]
The information score of a pruned graph $\gG_k$ with respect to task $\mY$ measures the fraction of task-relevant information retained over the entire spectrum, \ie
\begin{equation}
\label{eq:info-score}
    \gI(\gG_k) \coloneqq \frac{\widehat{I}_{q_\phi}(\gG_k;\mY) - \widehat{I}_{q_\phi}(\gG_K;\mY)}
    {\widehat{I}_{q_\phi}(\gG_0;\mY) - \widehat{I}_{q_\phi}(\gG_K;\mY)},
\end{equation}
where $\widehat{I}_{q_\phi}(\gG_k;\mY)$ is the estimated task-relevant MI defined in (\ref{eq:I_estimate}) with the parametric predictor $q_\phi(\cdot)$ (\eg GNNs).

\end{definition}
This score estimates the task-relevant MI retained in $\gG_k$ at the $k$th step. Like the complexity score, $\gI \in [0,1]$, with $\gI=1$ indicating full retention (same as the original graph) and $\gI=0$ indicating complete loss (same as the fully pruned graph).

The conceptual formulation of $\gI$ in (\ref{eq:info-score}) relates to $\widehat{I}_{q_\phi}(\gG;\mY)_\text{hold}$ closely in (\ref{eq:holdout_MI_est}). Following Proposition~\ref{prop:MI_est}, we employ $q_\phi(\cdot)$ to obtain the empirical estimation of (\ref{eq:info-score}), \ie
\begin{equation}
\label{eq:info-score-empirical}
    \gI(\gG_k) = \frac{\widehat{H}_{q_\phi}(\gG_k;\mY) - \widehat{H}_{q_\phi}(\gG_N;\mY)}
    {\widehat{H}_{q_\phi}(\gG_0;\mY) - \widehat{H}_{q_\phi}(\gG_N;\mY)}.
\end{equation}

\subsection{Global Measurements}
While $(\gI,\gC)$ provide a principled framework for evaluating graph simplification and information preservation during the pruning process, global measures are demanded for summarizing the entire pruning process and enabling comparison across methods. To this end, we formulate two global measurements. 

\paragraph{AUC-IC (Area Under the Information–Complexity Curve)}
The first metric aggregates the trade-off between $\gI$ and $\gC$ across all pruning steps. It is analogous to AUC-ROC, where performance is summarized by integrating the trade-off between true positive rate and false positive rate. AUC-IC has a minimum value of $0$, which corresponds to a pruning process that destroys information immediately. For the upper range, AUC-IC can exceed $1$ if pruning removes noisy edges that harm task performance, which is common in real-world graphs \citep{wang2024noisygl, fang2025quantifying}.

\paragraph{IBP (Information-Budget Point)}
The second metric captures the smallest complexity $\gC$ that achieves a target information retention threshold $\gI(\gG_k) \ge \tau$. 
IBP measures the smallest complexity \(\gC\) at which the pruned graph retains at least a target fraction of task-relevant information, \(\gI(\gG_k) \ge \tau\). Unlike AUC-IC, which evaluates the overall trade-off between information and sparsity, IBP emphasizes efficiency. It indicates how early a method can preserve sufficient information under limited complexity.

\section{\model: Gradient Boosting-motivated Graph Pruning}
\label{sec:model}
This section presents our multi-step graph pruning framework \model. Inspired by gradient boosting, \model~implements stepwise pruning guided by the empirical MI lower bound (Section~\ref{sec:MI_estimate}). Edges are ranked by their contribution to task loss, and the graph is simplified iteratively. We now detail the gradient boosting analogy and define the edge importance measures driving pruning.

\subsection{Motivation from Gradient Boosting}
The pruning problem can be interpreted through the lens of gradient boosting. In classical boosting, weak learners are sequentially fitted to the residuals of the current model to reduce the overall loss:  
\begin{equation}
f_k(x) = f_{k-1}(x) + \nu h_k(x),
\end{equation}
where $f_{k-1}(x)$ is the model at the previous iteration, $\nu$ is the learning rate, and $h_k(x)\in \gH$ is the weak learner chosen from a family of functions $\gH$ to approximate the negative gradient of a loss function $\gL$ with respect to the current predictions:
\begin{equation}
    h_k(x)=\argmin_{h\in\gH} \sum_{i=1}^n \big(-\frac{\partial \gL(y_i, f_{k-1}(x_i))}{\partial f_{k-1}(x_i)} - h(x_i)\big)^2.
\end{equation}

In graph pruning, we treat edges $e\in \gE$ analogously to weak learners. The objective is to iteratively remove edges that minimally contribute to reducing the task loss. At each pruning step $k$, the residual of an edge corresponds to its impact on the current model loss, indicating which edges can be safely removed without significantly degrading task performance. This perspective aligns naturally with the goal of maximizing task-relevant MI while minimizing graph complexity.

\subsection{Edge Importance and Stepwise Pruning}
The analogy to gradient boosting becomes clearer when viewing the adjacency matrix $\mA$ as the optimization target. For simplicity, we omit the subscript $k$ and discuss the general stepwise formulation. In gradient boosting, model parameters are updated along the negative loss gradient. Similarly, in \model, the gradient of the task loss $\gL$ with respect to $\mA$ indicates each edge’s effect on the prediction, with ${\partial \gL}/{\partial \mA_{ij}}$ serving as an edge-importance indicator.

Let $\gL(\mA,\theta)$ denote the task loss on graph $\gG_k$ with model $f_\theta(\cdot)$ parameterized by $\theta$. Edges with small gradient magnitudes contribute little to the task and can be removed. Following this principle, \model~prunes the graph in steps, repeatedly removing edges with minimal impact while keeping the most important connections. The residual contribution of edge $e_{ij}$ is:
\begin{equation}
\label{eq:importance-drop}
    S(e_{ij}) = \gL_{\rm val}(\mA \setminus \{e_{ij}\}, \theta) - \gL_{\rm val}(\mA, \theta),
\end{equation}
where $\mA \setminus \{e_{ij}\}$ denotes the adjacency matrix without $e_{ij}$. $\gL_{\rm val}$ on the validation set provides an unbiased estimate of $e_{ij}$’s impact.

Exact computation requires re-evaluating the model for each edge. To obtain an efficient approximation, $\mA$ is relaxed into a differentiable form, which estimates the edge importance as
\begin{equation}
\label{eq:importance-grad}
    S(e) = \frac{\partial \gL(f(\mA_k), \mY)}{\partial \mA_{k}(i, j)},
\end{equation}
where $\mA_{ij}$ corresponds to $e_{ij}$.

When optimizing the pruning process, edges with the lowest importance scores $S(e)$ are removed in batches, such that all $M$ edges are evenly pruned over multiple steps. The process continues until the graph is fully simplified. Algorithm~\ref{alg:igprune} summarizes \model. At each step, edge importance is computed using the model trained on the previously pruned graph, and the least important edges are removed to produce the next graph. 

The overall process integrates naturally with GNNs. The pruned graph is input to a GNN, and gradients from the task loss provide empirical signals for pruning, linking the information-theoretic objective, gradient-based optimization, and GNN inductive bias.

\begin{algorithm}[t]
\caption{Information-Guided Graph Pruning (\model)}
\label{alg:igprune}
\begin{algorithmic}[1]
\Require Graph $\gG_0=(V,E_0,X)$, labels $Y$, train/val masks; pruning steps $K$; 
         downstream epochs $T$; removal budget per step $\Delta e=\lfloor|E_0|/K\rfloor$
\Ensure Sequence of sparsified graphs $\{\gG_0,\gG_1,\dots,\gG_K\}$
\vspace{0.2em}

\State Initialize $E_{\text{cur}}\!\leftarrow\!E_0$, $\gG\!\leftarrow\!\{\gG_0\}$
\For{$k=1$ to $K$}
    \If{$|E_{\text{cur}}|=0$}
        \State Append $(V,\varnothing,X)$ to $\gG$ and \textbf{continue}
    \EndIf
    \State $n_{\text{remove}} \leftarrow \min(\Delta e, |E_{\text{cur}}|)$
    \State Train a GNN $f_\theta$ on $\gG_k=(V,E_{\text{cur}},X)$ for $T$ epochs by minimizing
           $\mathcal{L}_{\text{train}}=\text{NLL}(f_\theta(X,E_{\text{cur}})[\text{train}],Y[\text{train}])$
    \State Evaluate baseline loss $\mathcal{L}_{\text{base}}$ on validation nodes
    \For{each edge $e_l\!\in\!E_{\text{cur}}$}
        \State Form $E_{-l}=E_{\text{cur}}\setminus\{e_l\}$ and evaluate loss 
               $\mathcal{L}_l$ on validation data
        \State Compute edge importance $S(e_l)=\mathcal{L}_l-\mathcal{L}_{\text{base}}$
    \EndFor
    \State Remove $n_{\text{remove}}$ edges with smallest $S(e_l)$
           \Comment{Least important edges are pruned}
    \State Update $E_{\text{cur}}$ and construct $\gG_k=(V,E_{\text{cur}},X)$
    \State Append $\gG_k$ to $\gG$
\EndFor
\State \Return $\gG=\{\gG_0,\gG_1,\dots,\gG_K\}$
\end{algorithmic}
\end{algorithm}

\subsection{Objective Formulation}
\begin{definition}[Task-Relevant Graph Pruning Objective]
\label{def:task_obj}
The goal of multi-step graph pruning is to simplify a graph $\gG$ by progressively removing edges while preserving information relevant to a downstream task.  
Formally, given an empirical data distribution $\gD$ over graphs and task labels $(\gG, \mathbf{Y}) \sim \gD$, the objective at step $k$ is
\begin{equation}\label{eq:obj}
\begin{aligned}
&\min_{\gP_k}\ 
\E_{(\gG, \mathbf{Y}) \sim \gD}
\big[\,\mathcal{L}\big(f(\gG_k), \mathbf{Y}\big)\,\big] \\
\text{s.t. } \quad &
\gG_k = \gP_k(\gG_{k-1}), \quad
|\gE_k| \ge (1 - {\tau_k}/{E}) |\gE_{k-1}|,
\end{aligned}
\end{equation}
where $f(\cdot)$ denotes the predictor, $\mathcal{L}(\cdot)$ the empirical loss function, and $\tau_k$ the pruning budget. 
\end{definition}
In addition to preserving task-relevant information, (\ref{eq:obj}) can be viewed from a second perspective: for a given level of preserved information, pruning also aims to minimize graph complexity. That is, among subgraphs with similar predictive power, $\gP_k$ selects those with fewer edges or simpler structure, thus effectively balancing information retention and structural simplicity.

\section{Empirical Analysis}
This section presents empirical evaluations of \model. We aim to answer three key questions:
\begin{enumerate}[label=\textbf{Q\arabic*},leftmargin=*]
    \item Does \model~effectively preserve task-relevant information during pruning? 
    \item How does \model~compare with existing baselines across diverse datasets?
    \item Can \model~provide new insights in real-world networks?
\end{enumerate}
We will answer the three questions accordingly by A1, A2, A3-1, and A3-2 in the result analysis sections. We begin by describing the experimental setup in detail.

\begin{table*}[t]
\centering
\caption{Performance comparison on the \textit{original label} task. Results are reported as mean$\pm$std over $5$ repetitions.}
\label{tab:main_part1}
\resizebox{0.7\textwidth}{!}{%
\begin{tabular}{lcccccccc}
\toprule
\multirow{2}{*}{Method} 
 & \multicolumn{2}{c}{\textbf{Cora}} 
 & \multicolumn{2}{c}{\textbf{CiteSeer}} 
 & \multicolumn{2}{c}{\textbf{PubMed}} 
 & \multicolumn{2}{c}{\textbf{Karate Club}} \\ 
\cmidrule(lr){2-3}\cmidrule(lr){4-5}\cmidrule(lr){6-7}\cmidrule(lr){8-9}
 & AUC-IC $\uparrow$ & IBP $\downarrow$ 
   & AUC-IC $\uparrow$ & IBP $\downarrow$ 
   & AUC-IC $\uparrow$ & IBP $\downarrow$ 
   & AUC-IC $\uparrow$ & IBP $\downarrow$ \\ 
\midrule
RE  & 0.67$\pm$0.02 & 0.6$\pm$0.0 & 0.84$\pm$0.12 & 0.4$\pm$0.1 & 0.57$\pm$0.03 & 0.6$\pm$0.1 & 0.68$\pm$0.03 & 0.4$\pm$0.1 \\
RN  & 0.59$\pm$0.02 & 0.7$\pm$0.0 & 0.76$\pm$0.13 & 0.6$\pm$0.1 & 0.49$\pm$0.02 & 0.6$\pm$0.1 & 0.55$\pm$0.01 & 0.8$\pm$0.0 \\
EFF & 0.56$\pm$0.02 & 0.7$\pm$0.0 & 0.76$\pm$0.12 & 0.7$\pm$0.1 & 0.69$\pm$0.02 & 0.5$\pm$0.1 & 0.71$\pm$0.01 & 0.5$\pm$0.0 \\
LD  & 0.46$\pm$0.02 & 0.8$\pm$0.1 & 0.41$\pm$0.07 & 0.8$\pm$0.1 & 0.65$\pm$0.02 & 0.5$\pm$0.0 & 0.43$\pm$0.01 & 0.9$\pm$0.0 \\
LS  & 0.60$\pm$0.02 & 0.5$\pm$0.0 & 0.49$\pm$0.08 & 0.7$\pm$0.1 & 0.65$\pm$0.02 & 0.5$\pm$0.0 & 0.92$\pm$0.01 & 0.3$\pm$0.0 \\
SCAN & 0.72$\pm$0.02 & 0.5$\pm$0.0 & 0.87$\pm$0.13 & 0.5$\pm$0.1 & 0.56$\pm$0.02 & 0.7$\pm$0.0 & 0.86$\pm$0.01 & 0.4$\pm$0.0 \\
SO   & 0.49$\pm$0.01 & 1.0$\pm$0.0 & 0.59$\pm$0.08 & 0.7$\pm$0.1 & 0.16$\pm$0.02 & 1.0$\pm$0.0 & 0.69$\pm$0.01 & 0.5$\pm$0.0 \\
\textsc{PRI-Graphs} & 0.52$\pm$0.02 & 0.6$\pm$0.0 & 0.58$\pm$0.10 & 0.7$\pm$0.1 & / & / & 0.57$\pm$0.01 & 0.6$\pm$0.0 \\
\model & 1.12$\pm$0.03 & 0.3$\pm$0.0 & 1.70$\pm$0.29 & 0.2$\pm$0.1 & 0.66$\pm$0.02 & 0.5$\pm$0.1 & 0.78$\pm$0.01 & 0.3$\pm$0.0 \\
\bottomrule
\end{tabular}%
}
\end{table*}

\begin{figure*}[htbp]
    \centering
    \begin{minipage}[b]{0.485\linewidth}
        \centering
        \includegraphics[width=\linewidth]{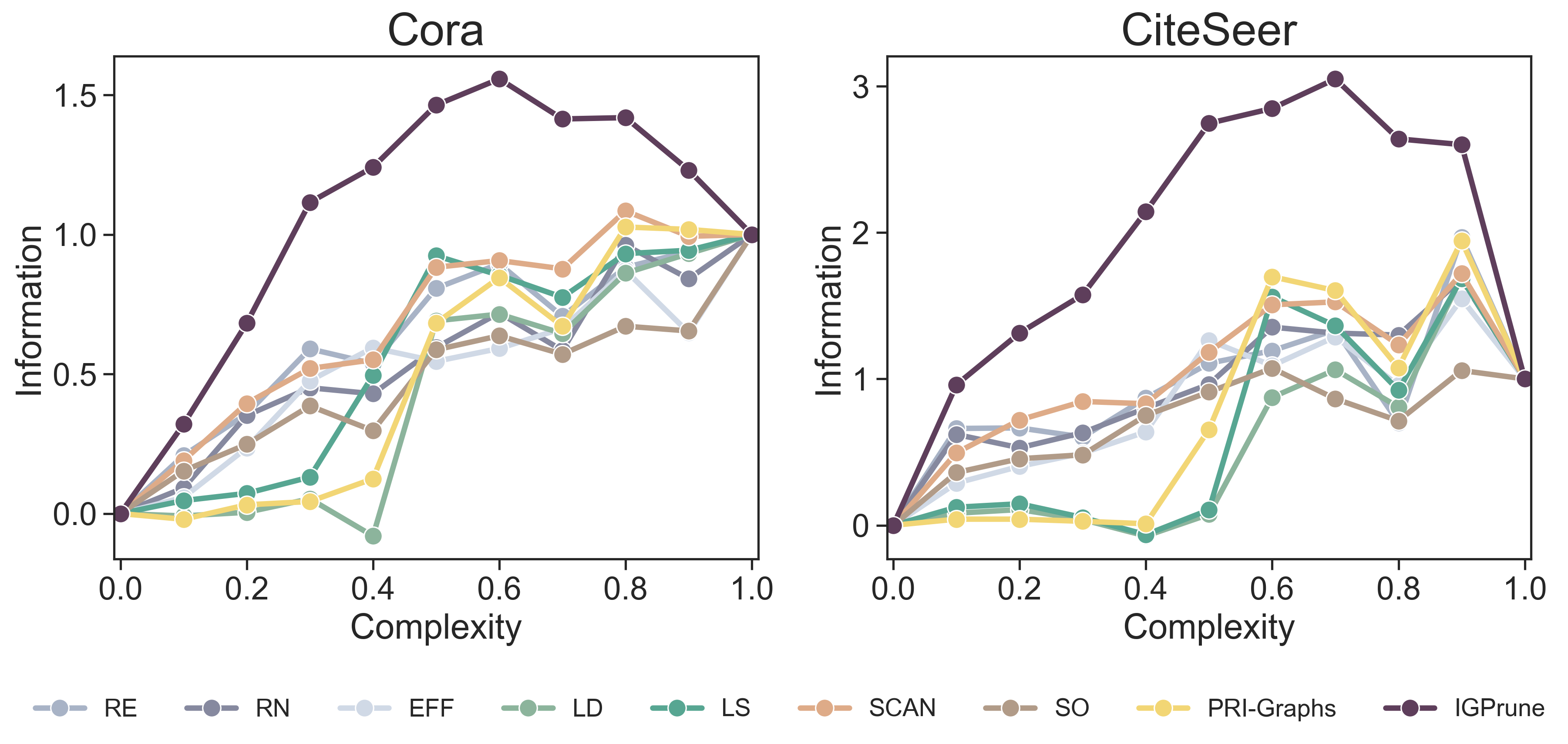}
        \caption{IC curve in the original label task.}
        \label{fig:ic_curve}
    \end{minipage}
    \hfill
    \begin{minipage}[b]{0.485\linewidth}
        \centering
        \includegraphics[width=\linewidth]{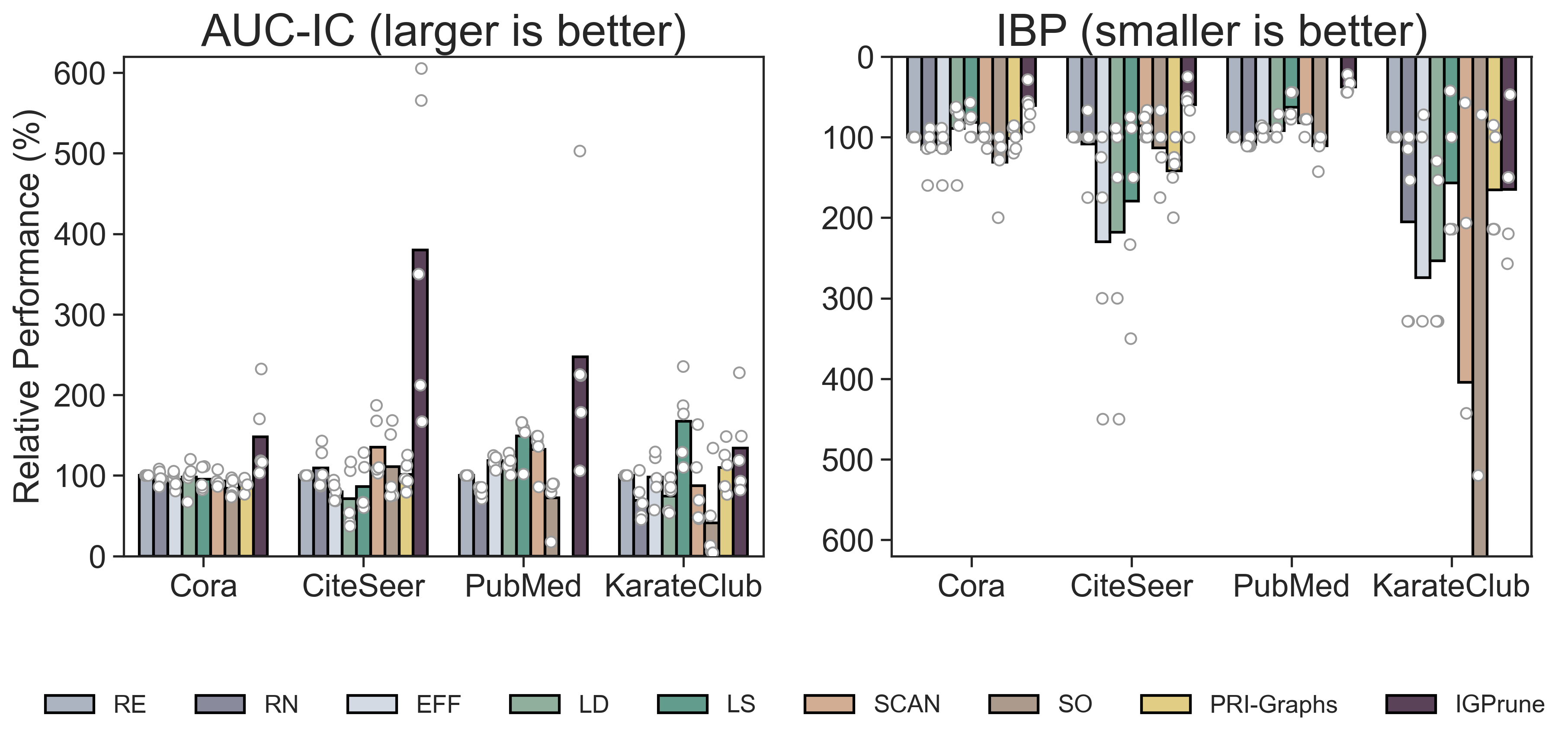}
        \caption{Summary of relative performance on all tasks.}
        \label{fig:bar_scatter}
    \end{minipage}
\end{figure*}

\subsection{Experimental Setup}
\paragraph{Datasets} 
Our evaluation covers three categories. (1) Three citation networks \citep{bojchevski2017deep}: \textbf{Cora}, \textbf{Citeseer}, and \textbf{PubMed}; (2) one social networks: \textbf{Karate Club} \citep{zachary1977information}; (3) two metabolic networks: \textbf{ME} and \textbf{MT} \citep{liu2022comparison,zhou2024odnet}. These datasets collectively represent a range of structural and semantic complexities, enabling comprehensive assessment across classic benchmarks and real-world scientific networks.

\paragraph{Downstream Tasks} 
As defined in Definition~\ref{def:task_obj}, graph pruning can be formulated with respect to different downstream objectives. For the quantitative analysis on citation and social networks, we define five node classification tasks to evaluate how each model performs under distinct structural or semantic criteria, including \textbf{Original label}, \textbf{Closeness centrality}, \textbf{Degree centrality}, \textbf{Degree}, and \textbf{PageRank}. More details are in Appendix~\ref{sec:app:tasks}. For \textbf{Cora}, \textbf{CiteSeer}, and \textbf{PubMed}, we use the \texttt{Planetoid} splits in \texttt{PyTorch Geometric}, following \citet{kipf2016gcn}.
For synthetic tasks and datasets without predefined splits, \ie \textbf{KarateClub}, \textbf{ME}, and \textbf{MT}, we randomly partition the data into $60\%:20\%:20\%$ training, validation, and test sets with a fixed random seed ($42$) for reproducibility.

\paragraph{Baseline Models}
\model~is compared against diverse pruning strategies covering random pruning strategies, structural heuristics, and information-theoretic approach, including Random Edge (RE), Random Node (RN), Edge Forest Fire (EFF), Local Degree (LD), Local Similarity (LS), SCAN Structural Similarity (SCAN), Simmelian Overlap (SO), and  \textsc{PRI-Graphs} \citep{yu2022principle}. The details are in Appendix~\ref{sec:app:baseline}.

\paragraph{Model Architecture} 
All baseline methods are evaluated under a unified $K$-step pruning setting with $K=10$. For methods requiring auxiliary models to compute edge scores on updated graphs, we employ a simple $2$-layer GCNs \citep{kipf2016gcn} from \texttt{PyTorch\_Geometric v2.7.0} \citep{Fey_et_al_2025, Fey/Lenssen/2019}. Each GCN layer has a hidden size of $128$, connected with ReLU activations. On all baselines and benchmarks, the GCNs are trained with random initialization on each simplified graph with a learning rate of $10^{-2}$, weight decay of $5\times 10^{-4}$. All experiments are conducted on a NVIDIA® GeForce RTX™ 4090.
The implementation is at \url{https://anonymous.4open.science/r/IGPrune-2D1B}.

\paragraph{Evaluation metrics}
The quality of graph pruning is evaluated using multiple criteria. Following the metrics established in Section~\ref{sec:IC_score}, we primarily adopt two global indicators: AUC-IC (higher is better) and IBP (lower is better), which jointly capture the trade-off between pruning complexity and information preservation. Additionally, we further examine the pruning behavior of each model by visualizing the IC curve and inspecting the pruned graphs at several key pruning steps. Notably, IBP is defined as the minimum complexity required to retain $\delta = 0.8$ of the original information.

\subsection{Quantitative Results and Analysis}
We first evaluate all baselines on four classic open benchmarks across five tasks on AUC-IC and IBP scores (Table~\ref{tab:main_part1}, Figure~\ref{fig:bar_scatter} and Tables~\ref{tab:main_part2} in the Appendix) and visualizing the IC curves (Figure~\ref{fig:ic_curve}) to address the first two questions.

\begin{figure*}[t]
    \centering
    \includegraphics[width=\linewidth]{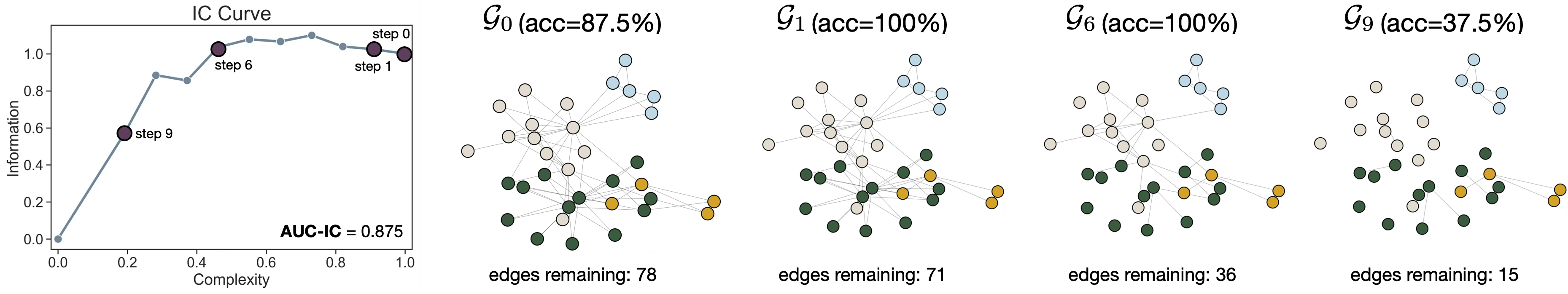}
    \caption{Pruning process on the Karate Club network. Node colors denote ground-truth community labels.}
    \label{fig:karate}
\end{figure*}

\paragraph{A1. Preservation of Task-Relevant Information}
To address Q1 regarding whether \model~effectively preserves task-relevant information during pruning, we first note that the task-relevant \emph{Information Measure} is computed on the downstream test set via the NLL loss of the predictor. 
Consequently, changes in this information closely reflect variations in classification performance. 
We then examine \model's performance across different benchmarks, tasks, and evaluation metrics. 
As shown in {Table~\ref{tab:main_part1}, Figure~\ref{fig:bar_scatter} and Table~\ref{tab:main_part2}}, \model~consistently achieves high AUC-IC scores across almost all tasks and datasets, indicating that essential node-level semantic and structural information is preserved throughout the pruning trajectory. 
In some cases (\eg, the original label task on \textbf{Cora} and \textbf{Citeseer}), \model~even obtains scores exceeding 1, suggesting that the model initially removes noisy edges that mislead label prediction. 
The corresponding IC curves are presented in Figure~\ref{fig:ic_curve}. 
In the early stages of pruning, the information score (y-axis) sometimes exceeds that of the original graph (when complexity = 1), indicating that removed edges are likely misleading, and excluding them from the graph could improve downstream performance. 
For instance, on \textbf{Cora}, the information of the original graph is $1.0$, while at complexity = $0.6$ (corresponding to $40\%$ of edges removed), it reaches its peak at $1.5$. 
This observation demonstrates that \model~not only preserves task-relevant information but also removes task-irrelevant edges.  
Moreover, in both cases, the sharp drop in information occurs only at later pruning stages. 
For example, on \textbf{Citeseer}, the measured information remains above that of the raw graph until only $10\%$ of edges remain. 
This further illustrates that \model~progressively retains the most important edges for the task throughout the pruning process.

\paragraph{A2. Comparison with Baseline Methods}

We further compare \model\ with baseline methods. 
Figure~\ref{fig:bar_scatter} presents a summary of the relative performance across all datasets and tasks, measured as the ratio with respect to the \textsc{RE} baseline. For each dataset, five downstream tasks, including one with original labels and four with synthetic labels, are evaluated. The bar heights represent the average performance across these tasks, while the white dots indicate individual task results. Higher AUC-IC values correspond to better information–complexity trade-offs, whereas lower IBP values indicate more efficient pruning. To improve visualization, excessively large IBP values are truncated to emphasize the comparative trends among methods.
Table~\ref{tab:main_part1} reports the performance of each model on the original label task. Performance on the other four tasks is provided in Table~\ref{tab:main_part2} in Appendix.

Overall, \model~consistently achieves the top performance across datasets and tasks. Compared to heuristic pruning methods (EFF, LD, and LS), \model~has higher AUC-IC on average and reaches lower information thresholds. Randomized methods (RE, RN) show unstable performance and degrade substantially on larger networks, while methods based on local similarity or overlap (SCAN, SO) have limited effectiveness on centrality-based tasks, reflecting their inability to capture global dependencies. Across datasets, \model~maintains robust performance as graph size and sparsity increase (see Table~\ref{tab:stats:node_classification} in Appendix). On smaller citation networks (\textbf{Cora}, \textbf{Citeseer}), both \model~and \textsc{PRI-Graphs} perform well, but \model~achieves better trade-offs between complexity and information preservation with lower computational cost. On the larger \textbf{PubMed} graph, \textsc{PRI-Graphs} fails to scale and times out, whereas \model~remains efficient and achieves the highest AUC-IC.
Task difficulty appears to vary across classification objectives. \model~consistently maintains high information retention during pruning, suggesting its ability to preserve essential structures at multiple scales. Nonetheless, \model’s strong performance on all five tasks demonstrates its ability to preserve essential structures throughout the pruning process, validating the effectiveness of its information–complexity optimization principle.

\subsection{Case Study on Karate Club Network: Demonstration of Pruning Dynamics}

To illustrate how \model~preserves the core structure of social networks during multi-step pruning, we conduct a case study on the \textit{Karate Club} dataset. We adopt the same setup as before and define the task as the original 4-class node classification with $K=10$. The multi-step pruning process is shown in Figure~\ref{fig:karate}, displaying the IC curve and network visualization at four key steps.

\paragraph{A3-1. Insights from Multi-Step Pruning}
The Karate Club network starts with $34$ nodes and $78$ undirected edges (step $0$). After step $6$, only $36$ edges remain, which is roughly half of the original connections. Despite this substantial sparsification, the downstream GCN classifier maintains perfect $100\%$ accuracy on the original node labels, indicating that task-relevant structural information is effectively preserved. Notably, a slight improvement in classification accuracy is observed from the first pruning step, where most of them are inter-community connections. This implies that \textit{early pruning removes noisy or cross-group links that could be misleading to the task}. A closer examination of the simplified graph at step $6$ shows that most remaining edges connect nodes sharing the same label, forming cohesive intra-community subgraphs. This pattern aligns with the semantics of social networks, where node labels are primarily determined by within-group connectivity rather than global link density. \model~carefully maintains overall connectivity, ensuring that message-passing in the GCN remains effective. Even after substantial edge reduction, nodes remain largely within a single connected component, highlighting that \model~\textit{prunes edges adaptively while preserving critical links that sustain information flow}. Furthermore, when pruning continues beyond a critical point (\eg fewer than $15$ edges after step $9$), network connectivity is severely disrupted, leading to a sharp drop in classification accuracy. This observation confirms that \textit{excessive removal of inter-community or bridging edges impairs information propagation in GNNs}. 

Overall, this case study demonstrates that \model~achieves a balanced trade-off between sparsity and information retention, preserving the structural essence and connectivity of the social network until pruning becomes overly aggressive.

\begin{figure*}[t]
    \includegraphics[width=0.95\textwidth]{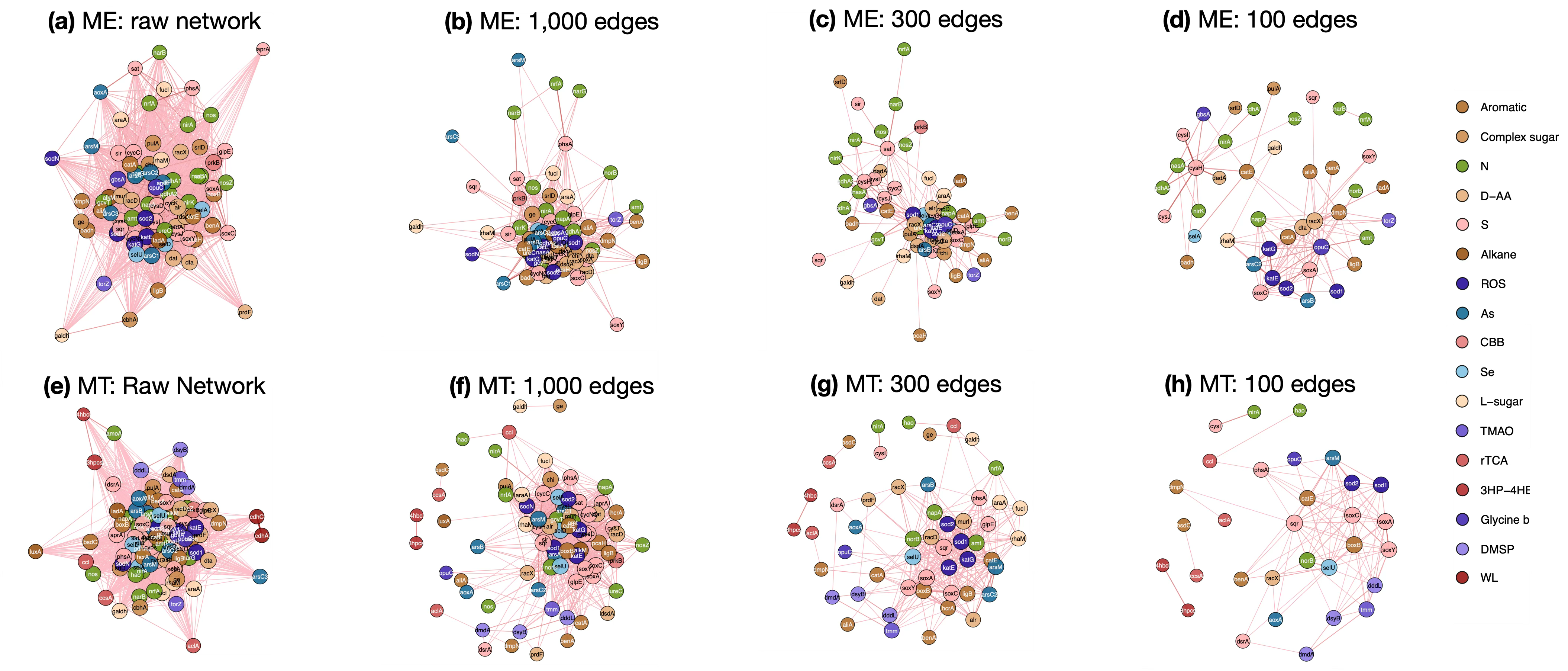}
    \caption{Multi-step pruning by \model~on two microbial gene co-occurrence networks from Mount Everest (ME) and the Mariana Trench (MT). Each subplot shows network snapshots with $2,000+$ (raw), $1,000$, $300$, and $100$ edges.}
    \label{fig:memt}
\end{figure*}

\subsection{Case Study on Gene Co-occurrence Networks: Discovery of Microbial Adaptation in Extreme Environments}
\label{sec:case_study}
To further evaluate whether \model~can reveal new insights from real-world networks, we analyze microbial gene co-occurrence networks from two extreme environments, \ie Mount Everest (ME) and the Mariana Trench (MT) \citep{liu2022comparison}. This case study tests \model’s capability to extract interpretable organization patterns from densely connected biological graphs, moving beyond benchmark demonstrations to scientific discovery.

Microorganisms are among the most abundant and diverse forms of life on Earth. They inhabit from deep-sea trenches to high-altitude plateaus \citep{shu2022microbial}. Their diversity arises from distinct metabolic pathways, enabling energy utilization and stress tolerance across extreme physical and chemical conditions. Comparing microbial metabolic systems across environments helps explain how global biogeochemical cycles operate and how ecosystems maintain their stability. It also provides insights for applied research areas such as bioremediation and industrial biotechnology \citep{louca2018function, coelho2022towards}. Metabolic gene networks provide a computational framework for examining these organizational principles. However, identifying functional modules remains challenging due to their large size and redundancy.

\paragraph{A3-2. Core Structural Divergence Between Extreme Environments}
The ME and MT represent two biological systems adapted to opposite environmental extremes. In both networks, nodes denote functional genes involved in carbon, nitrogen, sulfur, or other metabolic pathways. Edges connect co-occurred genes within the same microbial species, with weights indicating co-occurrence frequency. We set the pathway label prediction task and kept the basic model configuration unchanged, except for using $K = 100$ for finer resolution. Figure~\ref{fig:memt} visualizes the multi-step pruning results, showing graphs with $2,000+$ (original), $1,000$, $300$, and $100$ edges, respectively. The dense original networks obscure structural interpretation, whereas \model~progressively extracts stable core architectures that persist through pruning. Particularly, \textbf{ME shows a bipartite modularity and MT forms a core-periphery organization}. 
(1) ME includes two stable modules. Module one links sulfur (S) and nitrogen (N) metabolism genes, which suggests tight nutrient-cycling integration. Module two forms a highly interconnected cluster of three distinct functional categories, \ie reactive oxygen species (ROS) resistance, heavy metal detoxification (notably arsenic, As), and complex carbon degradation (\eg aromatic compounds, D-amino acids). This pattern reveals key genetic adaptations to UV-induced oxidative stress at high altitude. The co-clustering within module two even after aggressive pruning (\eg Figure~\ref{fig:memt}(c) with $10\%$ edges remaining) implies frequent co-occurrence and functional interdependence. The two modules revealed by \model~indicates \textit{ME microbes face two parallel but equally critical selective pressures of environmental stress tolerance and nutrient acquisition.} 
(2) MT retains a centralized core–periphery structure across all pruning levels. The core integrates mechanisms for extreme hydrostatic pressure adaptation (\eg ROS resistance, glycine betaine, DMSP synthesis), sulfur oxidation (\eg chemosynthetic energy metabolism), and aromatic carbon utilization. The early removal of peripheral metabolic genes (\eg C, N, and S pathways) suggests a non-essential role of general-purpose metabolic functions. In contrast, the persistence of the integrated core module for pressure and energy adaptation under strong pruning indicates that \textit{these gene combinations are under consistent selective pressure and represent the fundamental metabolic organization strategy of trench microbial communities.}

\section{Conclusion and Discussion}
This study introduces \model, an information- and complexity-guided framework that progressively prunes redundant connections while preserving task-relevant and semantically meaningful structures for graphs. Across benchmark datasets and real-world networks, \model~maintains predictive performance under high sparsity and reveals interpretable organization principles beyond what heuristic methods capture.
These findings demonstrate that information-theoretic pruning can serve as both a robust algorithmic tool and a means of scientific discovery. The framework’s generality makes it applicable to web-scale, social, and biological graphs, offering a controllable and interpretable way to expose core structures in complex networks. Future extensions to dynamic, heterogeneous, or multimodal settings may further enhance its utility for understanding and modeling large real-world systems.


\section*{Acknowledgments}
This work was supported by the grants from National Science Foundation of China (Grant Number 92451301; 62302291), the National Key Research and Development Program of China (2024YFA0917603), and Computational Biology Key Program of Shanghai Science and Technology Commission (23JS1400600).

\bibliographystyle{ACM-Reference-Format}
\bibliography{2reference}

\appendix
\input{2appendix}

\end{document}

%% file: math_commands.tex
\usepackage{amsmath,amsfonts,bm}

\def\eqref#1{equation~\ref{#1}}

\def\1{\bm{1}}

\def\mA{{\bm{A}}}

\def\mC{{\bm{C}}}

\def\mI{{\bm{I}}}

\def\mX{{\bm{X}}}
\def\mY{{\bm{Y}}}

\DeclareMathAlphabet{\mathsfit}{\encodingdefault}{\sfdefault}{m}{sl}
\SetMathAlphabet{\mathsfit}{bold}{\encodingdefault}{\sfdefault}{bx}{n}

\def\gC{{\mathcal{C}}}
\def\gD{{\mathcal{D}}}
\def\gE{{\mathcal{E}}}

\def\gG{{\mathcal{G}}}
\def\gH{{\mathcal{H}}}
\def\gI{{\mathcal{I}}}

\def\gL{{\mathcal{L}}}

\def\gP{{\mathcal{P}}}

\def\gV{{\mathcal{V}}}

\newcommand{\E}{\mathbb{E}}

\newcommand{\R}{\mathbb{R}}

\newcommand{\KL}{D_{\mathrm{KL}}}

\DeclareMathOperator*{\argmin}{arg\,min}

%% file: 2appendix.tex
\newpage

\section{Proof of Proposition 1}
\label{app:proof}
\begin{proposition}[Predictor-Based Mutual Information Lower Bound]
\label{prop:MI_est}
Let $q_\phi(\mY\mid\gG)$ be a parametric predictor with learnable parameters $\phi$, trained on labeled samples $\{(\gG_i, y_i)\}_{i=1}^M$ to approximate $p(\mY\mid\gG)$. 
Then, the MI between the graph $\gG$ and the task target $\mY$ can be lower-bounded by the empirical negative log-likelihood (NLL) of the predictor:
\begin{equation}
\label{eq:I_estimate}
    \widehat{I}_{q_\phi}(\gG;\mY)
    \approx 
    \hat{H}(\mY)
    - \frac{1}{M}\sum_{i=1}^M [-\log q_\phi(y_i \mid \gG_i)],
\end{equation}
where $M$ denotes the number of labeled samples, $y_i$ is the task label of sample $i$, and $\hat{H}(\mY)$ is the empirical entropy of the target variable.
\end{proposition}

\begin{proof}
We start from the task-relevant formulation of MI by updating (\ref{eq:mi_def}), which gives 
\begin{equation}
\label{eq:mi_def_2}
    I(\gG;\mY) = H(\mY) - H(\mY\mid\gG).
\end{equation}
The true conditional distribution $P(\mY\mid\gG)$ is unknown. To obtain a tractable estimate, we replace it with the predictive distribution by the classifier $q_\phi(\mY \mid \gG)$ , which yields the cross-entropy
\begin{align}
\label{eq:ce1}
    \widehat{H}_{q_\phi}(\mY\mid\gG) &= \E_{p(\gG,\mY)}[-\log q_{\phi}(\mY\mid\gG)] \\
    &= H(\mY\mid\gG) + \E_{p(\gG)}\Big[\KL(p(\mY\mid\gG)\,\|\,q_\phi(\mY\mid\gG))\Big] \notag
\end{align}
with $\KL(\cdot)$ denotes KL divergence. Apply (\ref{eq:ce1}) to (\ref{eq:mi_def_2}) results in a task-based estimator, \ie
\begin{equation}
\begin{aligned}
\label{eq:mi_estimator1}
    \widehat{I}_{q_\phi}(\gG;\mY) &\coloneqq H(\mY) - \widehat{H}_{q_\phi}(\mY\mid\gG)\\
    &= H(\mY) - \E_{p(\gG,\mY)}[-\log q_\phi(\mY\mid\gG)].
\end{aligned}
\end{equation}

Given the non-negative nature of  $\KL(\cdot)$, it holds that
\[
\widehat{H}_{q_\phi}(\mY\mid\gG) \ge H(\mY\mid\gG),
\]
which implies 
\[
\widehat{I}_{q_\phi}(\gG;\mY) \le I(\gG;\mY).
\]
Thus, $\widehat{I}_{q_\phi}(\gG;\mY)$ provides a lower bound of the true MI $I(\gG;\mY)$. The gap corresponds to the expected KL divergence between the true conditional $p(\mY\mid\gG)$ and the classifier $q_\phi(\mY\mid\gG)$.

Empirically, $H(\mY)$ is approximated from the empirical label distribution $\hat{p}(\mY)$, \ie
\begin{equation}
\label{eq:H_estimate}
    \hat{H}(\mY) = -\sum_y \hat{p}(y) \log \hat{p}(y).
\end{equation}
By furthering replace the expectation over the joint distribution $\E_{p(\gG,\mY)}[-\log q_\phi(\mY\mid\gG)]$ with the NLL, (\ref{eq:mi_estimator1}) becomes
\begin{equation*}
    \widehat{I}_{q_\phi}(\gG;\mY)
    \approx
    \hat{H}(\mY)
    - \frac{1}{M}\sum_{i=1}^M [-\log q_\phi(y_i\mid\gG_i)].
\end{equation*}
Minimizing the empirical NLL thus approximately maximizes the lower bound of task-relevant MI, and it provides a practical objective for tracking information retention in graph pruning.
\end{proof}

\section{Statistics of Benchmark Dataset}
Table~\ref{tab:stats:node_classification} summarizes the key statistics of all datasets used for node classification.
\textbf{Cora}, \textbf{Citeseer}, and \textbf{PubMed} are standard citation benchmarks; \textbf{KarateClub} is a small social network; \textbf{ME} and \textbf{MT} are two microbial gene co-occurrence networks.

\begin{table*}[!htbp]
\caption{Statistical summary of benchmark datasets.}
\label{tab:stats:node_classification}
\begin{center}
    \begin{tabular}{lcccccc}
    \toprule
    & \textbf{Cora} & \textbf{Citeseer} & \textbf{PubMed} & \textbf{KarateClub} & \textbf{ME} & \textbf{MT} \\
    \midrule
    \# Nodes & $2,708$ & $3,327$ & $19,717$ & $34$ & $79$ & $96$ \\
    \# Edges & $5,429$ & $4,732$ & $44,338$ & $78$ & $2,106$ & $2,210$ \\
    \# Features & $1,433$ & $3,703$ & $500$ & $34$ & $79$ & $96$ \\
    \# Classes & $7$ & $6$ & $3$ & $4$ & $15$ & $17$ \\
    \# Training Nodes & $140$ & $120$ & $60$ & $20$ & $47$ & $57$ \\
    \# Validation Nodes & $500$ & $500$ & $500$ & $6$ & $15$ & $19$ \\
    \# Test Nodes & $1,000$ & $1,000$ & $1,000$ & $8$ & $17$ & $20$ \\
    Label Rate & $0.052$ & $0.036$ & $0.003$ & $0.588$ & $0.595$ & $0.594$ \\
    Feature Scale & $\{0,1\}$ & $\{0,1\}$ & $[0.000, 1.263]$ & $\{0,1\}$ & $\{0,1\}$ & $\{0,1\}$ \\
    \bottomrule
    \end{tabular}
\end{center}
\end{table*}

\begin{figure*}[t]
    \centering
    \includegraphics[trim={0cm 9.5cm 0cm 7cm},clip,width=\linewidth]{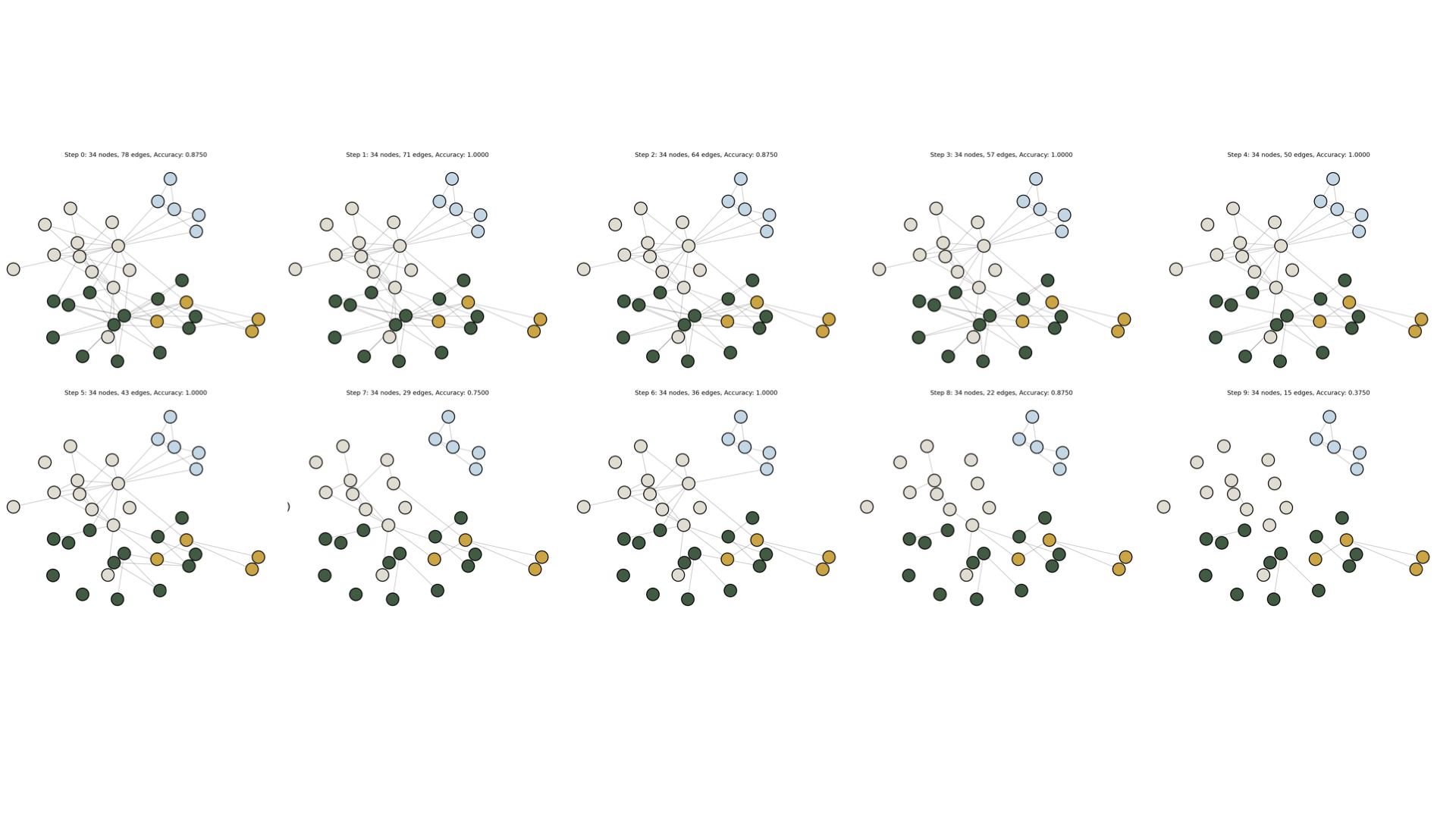}
    \caption{full visualizations of Pruning process on the Karate Club network.}
    \label{fig:karate}
\end{figure*}

\section{Additional Information on Experimental Setup}
\subsection{Evaluation Tasks}
\label{sec:app:tasks}
We give more details on the five tasks we evaluated in the quantitative analysis. 
\begin{enumerate}[leftmargin=*]
    \item \textbf{Original label}: Evaluates whether pruning preserves semantic node attributes. Each node is assigned its ground-truth label provided by the benchmark dataset (\eg research field in citation networks or community membership in social networks).
    \item \textbf{Closeness centrality}: Examines whether pruning retains essential shortest-path structures in the network. Nodes are grouped evenly into low, medium, and high according to their closeness centrality, which is calculated as the inverse of the average shortest-path distance to all other nodes.
    \item \textbf{Degree centrality}: Enables consistent comparison across graphs of different sizes. Labels are defined similarly to the degree classification but are based on normalized degree centrality.
    \item \textbf{Degree}: Measures the preservation of local connectivity. Nodes are evenly grouped into three categories (low, medium, and high) according to their degree quantiles.
    \item \textbf{PageRank}: Assesses whether pruning preserves global importance scores derived from random walks. Nodes are evenly divided into three categories (low, medium, and high) based on their PageRank values.
\end{enumerate}
For \textbf{Cora}, \textbf{CiteSeer}, and \textbf{PubMed}, we the \texttt{Planetoid} splits in \texttt{PyTorch Geometric}, following \citet{kipf2016gcn}.

\subsection{Baseline Methods}
\label{sec:app:baseline}
In quantitative analysis, we compare \model~with eight baseline methods. They are:
\begin{itemize}[leftmargin=*]
    \item Edge Forest Fire (EFF) \citep{angriman2023networkit}: a random-walk–based sparsification that samples edges according to a Forest-Fire process.
	\item Local Degree (LD): retains edges incident to high-degree nodes, prioritizing hub connectivity to preserve core local structure.
	\item Local Similarity (LS): keeps edges with high neighborhood similarity to preserve locally coherent links.
	\item Random Edge (RE): removes edges uniformly at random.
	\item Random Node (RN): samples a subset of nodes uniformly and deletes all incident edges.
	\item SCAN Structural Similarity (SCAN): detects clusters, hubs, and outliers using structural similarity and is used here by retaining intra-cluster edges to measure cluster-preserving pruning.
	\item Simmelian Overlap (SO): evaluates edge strength by higher-order overlap (shared strong ties) and retains edges with high Simmelian overlap.
    \item \textsc{PRI-Graphs} \citep{yu2022principle}: operates on the graph Laplacian to preserve key spectral and structural properties.
\end{itemize}
All baseline implementations are based on  \texttt{NetworKit} \citep{angriman2023networkit} available at \url{https://github.com/networkit/networkit}, except for \textsc{PRI-Graphs}, which is at \url{https://github.com/SJYuCNEL/PRI-Graphs}.

\begin{table}[t]
\centering
\caption{Time cost comparison (in seconds) between our method and \textsc{PRI-Graphs} across four datasets on the original classification task. Lower is better.}
\label{tab:timecost}
\begin{tabular}{lcc}
\toprule
\textbf{Dataset} & \textsc{PRI-Graph} & \model \\
\midrule
\textbf{Cora} & 1870.5 & \textbf{74.5} \\
\textbf{CiteSeer} & 2582.5 & \textbf{64.3} \\
\textbf{PubMed} & / & \textbf{644.4} \\
\textbf{Karate Club} & \textbf{10.2} & 11.6 \\
\bottomrule
\end{tabular}
\end{table}

\section{Additional Comparison with PRI-Graph.} 
We further compare PRI-Graph with \model~in terms of computational efficiency. This comparison is particularly relevant because both methods share similar theoretical motivations but differ in implementation complexity. As shown in Table~\ref{tab:timecost}, \model~achieves a shorter total runtime while maintaining superior pruning quality. The gap becomes more pronounced as graph size increases, where PRI-Graph also suffers from substantial memory consumption and fails to complete on the Pubmed dataset. This efficiency advantage mainly stems from the reliance of PRI-Graph on eigenvector decomposition and reconstruction operations, which introduce higher computational overhead compared to the lightweight optimization used in \model.

It is worth noting that traditional large-scale graph simplification methods such as \textit{networkit} can indeed scale to massive graphs by adopting local heuristics and edge sampling strategies. However, these approaches achieve scalability at the cost of global structural awareness. In contrast, \model~retains a principled global view of the graph’s information landscape, allowing it to identify globally optimal pruning directions rather than relying solely on local edge metrics. Consequently, while \textit{networkit}-style algorithms excel in scalability, \model~offers a more balanced trade-off between computational efficiency and global structure preservation, leading to consistently better pruning outcomes across both small and medium-sized graphs.

\section{Details of Karate Club Case Study}
This appendix provides full visualizations of the Karate Club case study, including all pruning steps, as a complement to the main text discussion.

\section{Results on Synthetic Label Tasks}\label{app:synthetic_results}
Table~\ref{tab:main_part2} reports the performance of all models on four synthetic label tasks (Closeness Centrality, Degree Centrality, Degree, and PageRank) across four datasets. We present results in terms of AUC-IC (↑) and IBP (↓). These results provide a comprehensive view of model behavior under different synthetic supervision settings.

\begin{table*}[t]
\centering
\caption{Results on task 1 under different metrics. 
We report AUC-IC (higher is better) and Info-threshold point (lower is better) across datasets. "/" indicates the method timed out on the dataset (more than 6 hours).}
\label{tab:main_part2}
\begin{tabular}{llcccccccc}
\toprule
\multirow{2}{*}{Task} & \multirow{2}{*}{Method} 
 & \multicolumn{2}{c}{\textbf{Cora}} 
 & \multicolumn{2}{c}{\textbf{CiteSeer}} 
 & \multicolumn{2}{c}{\textbf{PubMed}} 
 & \multicolumn{2}{c}{\textbf{Karate Club}} \\ 
\cmidrule(lr){3-4}\cmidrule(lr){5-6}\cmidrule(lr){7-8}\cmidrule(lr){9-10}
 & & AUC-IC $\uparrow$ & IBP $\downarrow$ 
   & AUC-IC $\uparrow$ & IBP $\downarrow$ 
   & AUC-IC $\uparrow$ & IBP $\downarrow$ 
   & AUC-IC $\uparrow$ & IBP $\downarrow$ \\ 
\midrule
\multirow{9}{*}{Closeness Centrality} 
 & RE & 0.62 & 0.7 & 0.76 & 0.4 & 0.41 & 0.9 & 0.77 & \textbf{0.1} \\
 & RN & 0.59 & 0.7 & 0.66 & 0.7 & 0.29 & 1.0 & 0.61 & \textbf{0.1} \\
 & EFF & 0.55 & 0.8 & 0.61 & 0.7 & 0.50 & 0.8 & 0.94 & 0.7 \\
 & LD & 0.65 & 0.5 & 0.81 & 0.4 & 0.52 & 0.8 & 0.75 & 0.3 \\
 & LS & 0.68 & 0.4 & 0.84 & 0.3 & 0.65 & 0.7 & \textbf{1.81} & 0.2 \\
 & SCAN & 0.57 & 0.7 & 0.78 & 0.4 & 0.58 & 0.7 & 0.85 & 0.4 \\
 & SO & 0.49 & 0.8 & 0.57 & 0.7 & 0.32 & 0.9 & 0.39 & 0.8 \\
 & \textsc{PRI-Graphs} & 0.59 & 0.6 & 0.71 & 0.5 & / & / & 1.14 & 0.2 \\
 & \model & \textbf{1.43} & \textbf{0.2} & \textbf{2.66} & \textbf{0.2} & \textbf{2.06} & \textbf{0.2} & 1.75 & 0.2 \\
\midrule
\multirow{9}{*}{Degree Centrality} 
 & RE & 0.56 & 0.9 & 0.85 & \textbf{0.2} & 0.34 & 0.9 & 0.81 & \textbf{0.1} \\
 & RN & 0.61 & 0.8 & 1.08 & \textbf{0.2} & 0.29 & 1.0 & 0.53 & 0.3 \\
 & EFF & 0.58 & 0.8 & 0.58 & 0.9 & 0.42 & 0.9 & 0.78 & \textbf{0.1} \\
 & LD & 0.57 & 0.6 & 0.31 & 0.9 & 0.38 & 0.9 & 0.66 & 0.3 \\
 & LS & 0.48 & 0.7 & 0.51 & 0.7 & 0.57 & \textbf{0.4} & \textbf{1.52} & 0.2 \\
 & SCAN & 0.50 & 0.8 & 1.42 & \textbf{0.2} & 0.51 & 0.7 & 0.56 & 0.6 \\
 & SO & 0.55 & 0.9 & 1.28 & \textbf{0.2} & 0.31 & 0.9 & 0.05 & 1.0 \\
 & \textsc{PRI-Graphs} & 0.52 & 0.8 & 0.95 & 0.4 & / & / & 1.02 & 0.2 \\
 & \model & \textbf{0.67} & \textbf{0.5} & \textbf{4.78} & 0.2 & \textbf{0.77} & 0.4 & 0.96 & 0.1 \\
\midrule
\multirow{9}{*}{Degree} 
 & RE & 0.58 & 0.7 & 0.76 & 0.3 & 0.34 & 0.9 & 1.18 & \textbf{0.1} \\
 & RN & 0.61 & 0.8 & 1.08 & \textbf{0.2} & 0.29 & 1.0 & 0.53 & 0.3 \\
 & EFF & 0.56 & 0.8 & 0.52 & 0.9 & 0.43 & 0.8 & 0.67 & 0.3 \\
 & LD & 0.57 & 0.6 & 0.31 & 0.9 & 0.38 & 0.9 & 0.66 & 0.3 \\
 & LS & 0.48 & 0.7 & 0.51 & 0.7 & 0.57 & 0.4 & \textbf{1.52} & 0.2 \\
 & SCAN & 0.50 & 0.8 & 1.42 & \textbf{0.2} & 0.51 & 0.7 & 0.56 & 0.6 \\
 & SO & 0.55 & 0.9 & 1.28 & \textbf{0.2} & 0.31 & 0.9 & 0.05 & 1.0 \\
 & \textsc{PRI-Graphs} & 0.52 & 0.8 & 0.95 & 0.4 & / & / & 1.02 & 0.2 \\
 & \model & \textbf{0.68} & \textbf{0.5} & \textbf{4.59} & 0.2 & \textbf{0.78} & \textbf{0.3} & 0.96 & 0.1 \\
\midrule
\multirow{9}{*}{PageRank} 
 & RE & 0.54 & 0.8 & 0.43 & 0.9 & 0.40 & 0.9 & 1.51 & \textbf{0.2} \\
 & RN & 0.52 & 0.9 & 0.38 & 0.9 & 0.32 & 1.0 & 0.76 & 0.3 \\
 & EFF & 0.57 & 0.8 & 0.41 & 0.9 & 0.47 & 0.9 & 1.29 & \textbf{0.2} \\
 & LD & \textbf{0.65} & \textbf{0.5} & 0.51 & 0.8 & 0.48 & 0.9 & 0.81 & 0.3 \\
 & LS & 0.59 & 0.6 & 0.56 & 0.8 & 0.62 & 0.7 & \textbf{1.66} & \textbf{0.2} \\
 & SCAN & 0.49 & 0.8 & 0.47 & 0.8 & 0.55 & 0.7 & 0.70 & 0.4 \\
 & SO & 0.40 & 0.9 & 0.37 & 0.9 & 0.35 & 1.0 & 0.20 & 1.0 \\
 & \textsc{PRI-Graphs} & 0.52 & 0.8 & 0.41 & 0.9 & / & / & 1.16 & \textbf{0.2} \\
 & \model & 0.55 & 0.7 & \textbf{0.72} & \textbf{0.5} & \textbf{0.72} & \textbf{0.4} & 1.41 & 0.4 \\
\bottomrule
\end{tabular}%
\end{table*}

It should be noted that \model~exhibits exceptionally high AUC-IC values on certain tasks (\eg Degree and Degree Centrality on Citeseer). This phenomenon arises because the initial prediction accuracy before pruning was relatively low, and the reported AUC-IC values are normalized with respect to the initial state, which can amplify the relative improvement magnitude.